\documentclass[twocolumn]{aastex63}

\newcommand{\etal}{et al.}

\newcommand\asca{{\it ASCA\/}}
\newcommand\chandra{{\it Chandra}}
\newcommand\xmm{{\it XMM-Newton\/}}

\newcommand\suzaku{{\it Suzaku\/}}
\newcommand\nustar{{\it NuSTAR\/}}
\newcommand\integral{{\it INTEGRAL\/}}
\newcommand\cco{CXOU~J171405.7$-$381031}
\newcommand\ccoshort{J171405}
\newcommand\ctb{CTB~37B}
\newcommand\snr{CTB~37B}
\newcommand\cxou{CXOU~J171405.7$-$381031}
\newcommand\cxoushort{J171405}
\newcommand\tev{HESS~J1713$-$381}

\newcommand\blob{XMMU~J171410.8$-$381442}

\def\simlt{\mathrel{\hbox{\rlap{\hbox{\lower4pt\hbox{$\sim$}}}\hbox{$<$}}}}
\def\simgt{\mathrel{\hbox{\rlap{\hbox{\lower4pt\hbox{$\sim$}}}\hbox{$>$}}}}

\shorttitle{Monitoring the Magnetar in SNR CTB 37B}
\shortauthors{Gotthelf \etal}

\begin{document}

\title{X-ray Monitoring of the Magnetar \cco\ in SNR CTB 37B}

\author{E. V. Gotthelf}
\altaffiliation{Columbia Astrophysics Laboratory, Columbia University, 550 West 120th Street, New York, NY 10027-6601, USA.}
\altaffiliation{Departament de F\'{\i}sica Qu\`antica i Astrof\'{\i}sica, Institut de Ci\`encies del Cosmos, Universitat de Barcelona, IEEC-UB, Mart\'{\i}\ i Franqu\`es 1, 08028, Barcelona, Spain.}

\author{J. P. Halpern}
\altaffiliation{Columbia Astrophysics Laboratory, Columbia University, 550 West 120th Street, New York, NY 10027-6601, USA.}

\author{K. Mori} 
\altaffiliation{Columbia Astrophysics Laboratory, Columbia University, 550 West 120th Street, New York, NY 10027-6601, USA.}

\author{A. M. Beloborodov}
\altaffiliation{Columbia Astrophysics Laboratory, Columbia University, 550 West 120th Street, New York, NY 10027-6601, USA.}

\correspondingauthor{E. V. Gotthelf; eric@astro.columbia.edu}


\begin{abstract} 

  We present the results of our 8~year X-ray monitoring campaign on 
  \cco, the magnetar associated with the faint supernova remnant (SNR) \ctb.
  It is among the youngest by inferred spin-down age, and most
  energetic in spin-down power of magnetars, and may contribute,
  at least partially, to the GeV and TeV emission coincident
  with the SNR.
  We use a series of \chandra, \xmm, and \nustar\ observations to
  characterize the timing and spectral properties of the magnetar.
  The spin-down rate of the pulsar almost doubled in $<1$ year and then
  decreased slowly to a more stable value.  Its X-ray flux varied by
  $\approx 50\%$, possibly correlated with the spin-down rate.
  The $1-79$~keV spectrum is well-characterized by an absorbed blackbody
  plus power-law model with an average temperature of $kT = 0.62 \pm
  0.04$~keV and photon index $\Gamma = 0.92 \pm 0.16$, or by a Comptonized
  blackbody with $kT = 0.55 \pm 0.04$~keV and an additional hard
  power law with  $\Gamma = 0.70 \pm 0.20$,
  In contrast with most magnetars, the pulsed signal is found to
  decrease with energy up to 6~keV, which is apparently caused by
  mixing with the hard spectral component that is pulse-phase shifted
  by $\approx0.43$ cycles from the soft X-rays.
  We also analyze the spectrum of the nearby, diffuse
  nonthermal source \blob, whose relation to the SNR is uncertain.
  
\end{abstract}

\keywords{ISM: individual (\ctb) --- pulsars: individual (\cco)
--- stars: neutron}

\section {Introduction}

Modeling the spectra of magnetars provide an important diagnostic for
understanding their emission mechanisms. Early literature characterized
magnetars as ``soft'' X-ray sources emitting in the
$<10$~keV X-ray band. Their spectra were typically fitted with
blackbody emission from a hot spot(s) on the neutron star (NS)
surface, plus a steep
power-law component with photon index $\Gamma \sim 4$,
possibly scattered from the magnetosphere (e.g., \citealt{mer95,van95}).
More recent \integral\ observations of the anomalous X-ray pulsar (AXP)
subclass of magnetars revealed a previously unrecognized,
flatter spectral component above $\sim$10~keV, extending up to
$>100$~keV, with a pulse modulation that increases with energy.
These hard X-rays, beginning with detections of 1E~1841$-$045 \citep{mol04}
1RXS J170849.0$-$400910 \citep{rev04}, and 4U~0142+61 \citep{den04},
are fitted with photon indices in the
range $\Gamma = 0.8-1.4$, much harder than an extrapolation of the
measured spectra below $<10$~keV.  The majority of the luminosity of
these objects is emitted above 10~keV in this hard spectral component.

\begin{deluxetable*}{lccccl}
\tablewidth{0pt}
\tablecaption{Log of X-ray Observations of \cco\ in \ctb}
\tablehead{
\colhead{Instrument/Mode\tablenotemark{a}} & \colhead{ObsID}  & \colhead{Date (UT)\tablenotemark{b}} & \colhead{Exposure (ks)\tablenotemark{c}} & \colhead{Epoch (MJD)\tablenotemark{d}} & Period (s)
}
\startdata
\asca\ GIS              &   54002030 & 1996 Sep 12 & 13.3 & 50338.9 & $3.7954(1)$ \\
\suzaku\ XIS            &  501007010 & 2007 Aug 27 & 82.8 & 53974.1 & \dots \\
\chandra\ ACIS-I/TE/VF  &       6692 & 2007 Feb 2 & 25.2 & 54133.5 & \dots \\
\chandra\ ACIS-S3/CC/F  &      10113 & 2009 Jan 25 & 30.1 & 54856.3 & $3.823056(17)$ \\
\chandra\ ACIS-S3/CC/F  &      11233 & 2010 Jan 30 & 30.1 & 55226.5 & $3.824936(18)$ \\
\xmm\ EPIC pn/FF        & 0606020101 & 2010 Mar 17 & 93.0 & 55273.2 & $3.825353(4)$ \\ 
\xmm\ EPIC pn/FF        & 0670330101 & 2012 Mar 13 & 11.5 & 55999.5 & $3.83039(4)  $ \\
\chandra\ ACIS-S3/CC/F  &      13749 & 2012 Jul 16 & 20.1 & 56124.3 & $3.831062(29)$ \\
\chandra\ ACIS-S3/CC/F  &      16762 & 2015 May 4 & 20.1 & 57146.6 & $3.835459(31)$ \\
\nustar\ FPM            &30001130002 & 2015 May 8 & 80.9 & 57151.1 & $3.8354346(69)$ \\
\chandra\ ACIS-S3/CC/F  &      16763 & 2015 Oct 13 & 19.2 & 57308.4 & $3.836025(38)$ \\ 
\nustar\ FPM            &30201031002 & 2016 Sep 22 & 78.6 & 57654.9 & $3.8375205(57)$ \\
\xmm\ EPIC pn/FF        & 0790870201 & 2016 Sep 23 & 27.5 & 57654.9 & $3.837494(18)$ \\
\xmm\ EPIC pn/FF        & 0790870301 & 2017 Feb 22 & 20.0 & 57806.5 & $3.837980(20)$
\enddata                                                     
\label{tab:logtable}
\tablenotetext{a}{Mission modes: \chandra\ Timed Exposure (TE), Continuous Clocking (CC), Faint Grading (F), Very Faint Grading (VF); \xmm\ Full Frame (FF).}
\tablenotetext{b}{Start date of observation.}
\tablenotetext{c}{Effective exposure time for the EPIC pn after time filtering.}  
\tablenotetext{d}{Epoch of period is the mid-time of the observation.}
\end{deluxetable*}

Hard X-rays from magnetars must be produced by nonthermal particles in
the magnetosphere.  A mechanism for this emission was recently
described by \citet{bel13a,bel13b}: electron-positron discharge creates
relativistic particles with Lorentz factors $\gamma\sim10^3$ near the
neutron star, the particles flow out along the extended magnetic field
lines and decelerate with $\gamma\propto B$, losing energy to resonant
scattering of thermal X-rays.  The spectrum emitted by the
decelerating plasma peaks above 100~keV, and its shape depends on the
angle of the rotation axis to the line of sight. The model
successfully reproduced the phase-resolved spectra of the
aforementioned magnetars, the three best-studied objects
\citep{has14}.  \nustar\ is well suited to studying the hard X-ray
spectral components of magnetars, up to 79~keV.
Its targets have included
SGR J1745$-$2900 \citep{mor13,kas14}, 
1E~1841$-$045 \citep{an13,an15},
1E~2259$+$586 \citep{vog14},
1E~1048.1$-$5937 \citep{an14,yan16},
4U~0142+61 \citep{ten15},
{ SGR 1806$-$20 \citep{you17a},
SGR J1935+2154 \citep{you17b}},
PSR~J1622$-$4950 \citep{cam18},
XTE J1810$-$197 \citep{got19},
and SGR 1900$+$14 \citep{tam19}.

\begin{figure}[]
\includegraphics[height=0.97\linewidth,angle=270]{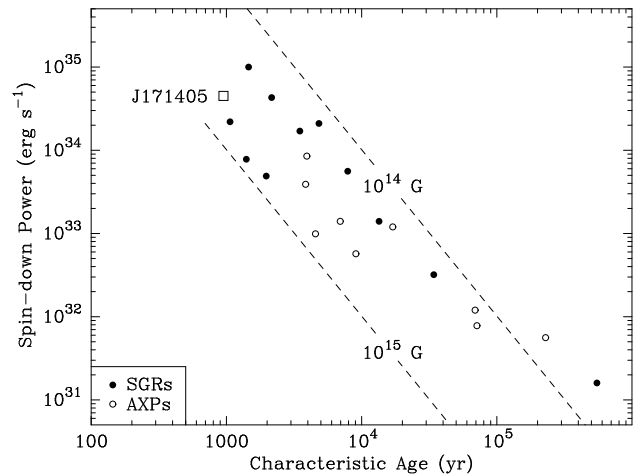}
\caption{
A summary of derived dipole spin-down properties of 21 magnetars based on
period and period-derivative data during relatively quiescent periods,
if available.  Two magnetars with very small or unmeasured spin-down
rates are omitted here.
}
\label{fig:comp}
\end{figure}

\citet{aha08a} discovered the \chandra\ point source \cxou\
in the supernova remnant (SNR) \ctb\ in a follow up effort to identify \tev,
a coincident TeV source \citep{aha06}. They considered the
\chandra\ source a candidate pulsar, albeit with an unusually soft,
non-thermal spectrum. \cite{Nakamura09} analyzed \chandra\ and
\suzaku\ spectra of \cxou\ (hereafter \cxoushort), suggesting that 
it is an AXP based on evidence of flux variability.
\citet[][Paper~I]{hal10a} discovered 3.82~s pulsations from \cxoushort\
that verifies this conjecture. \citet[][Paper~II]{hal10b} and \citet{sat10}
reported follow-up observations using \chandra\ and
\xmm, respectively, that measure the period derivative of the pulsar,
establishing its quantitative magnetar properties. \cxoushort\
has a higher spin-down power and younger characteristic
age than most magnetars (Figure~\ref{fig:comp}), falling among
the most energetic soft gamma-ray repeaters (SGRs).  Its spin-down
power is comparable to its X-ray luminosity. 

Although \tev\ coincides with the SNR, its TeV
structure has not been spatially resolved, and the young
age and rapid spin-down of \cxoushort\ suggest the possibility (Paper II)
that the pulsar contributes to the TeV emission via inverse Compton
scattering by a relic pulsar wind nebula (PWN).  Moreover, the
presence of the nearby unidentified hard, diffuse X-ray source \blob\ 
adds further uncertainty about the origin of the TeV emission.

As described in Section~\ref{sec:obs}
we have obtained new \chandra, \xmm, and \nustar\ observations of
\ccoshort\ in order to monitor its timing and spectrum.  A prime
motivation was to search for any precursor of an impending SGR outburst,
as may be anticipated from its location among the SGRs in Figure~\ref{fig:comp}.
A log of X-ray observations is presented in Table~\ref{tab:logtable}.
In Sections~\ref{sec:image}--\ref{sec:spectra} we describe the analysis
of these data sets, along
with available archival data spanning two decades. We show that the
pulsar spins down erratically, and find that
the \nustar\ X-ray observations clearly detect emission above 10~keV. 
In Section~\ref{sec:spectra} we also provide a spectral analysis of  
the nearby hard, diffuse nonthermal source.
Finally, we discuss our results in Section~\ref{sec:disc}.

\begin{figure*}[]
\hfill

\centerline{
\includegraphics[height=0.48\linewidth,angle=270]{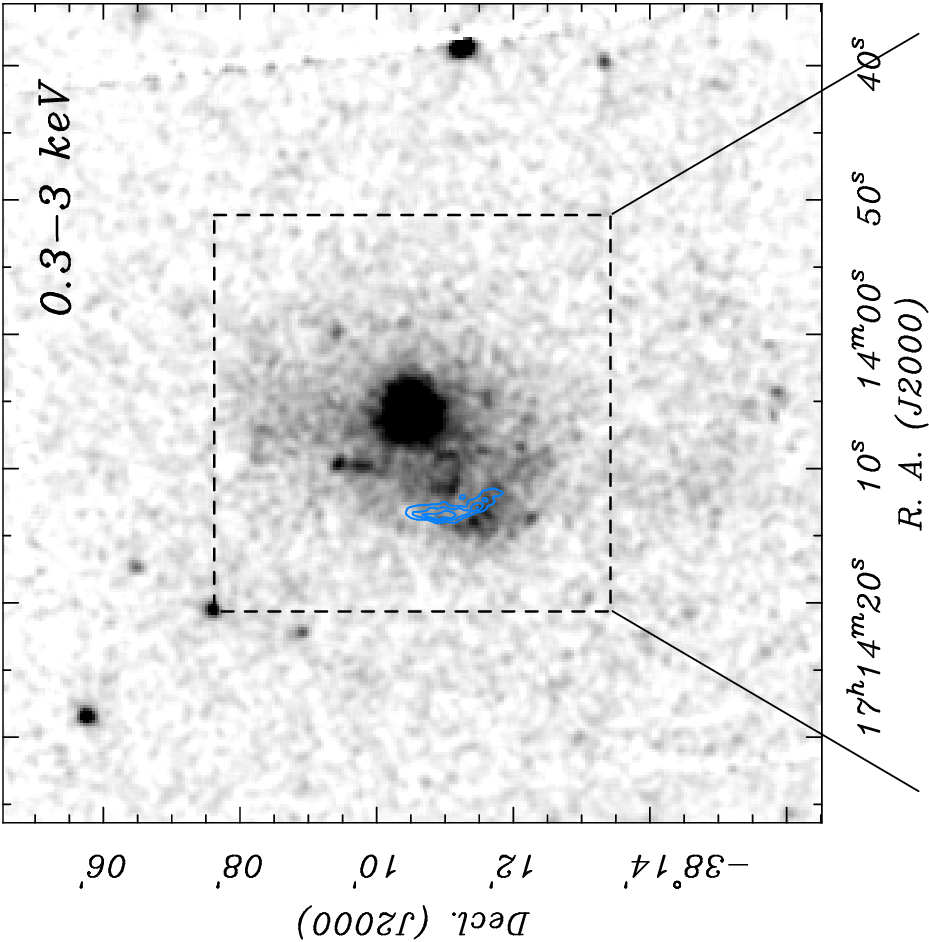}
\hfill
\includegraphics[height=0.48\linewidth,angle=270]{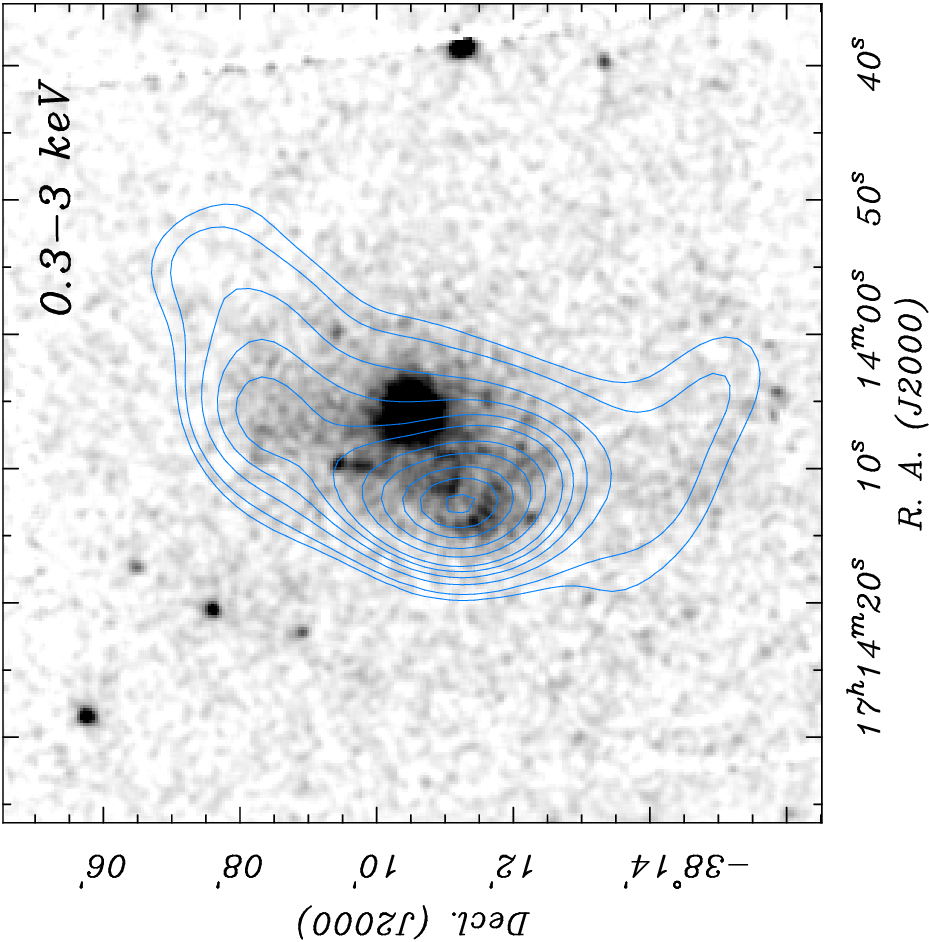}
}
\centerline{
\includegraphics[height=0.48\linewidth,angle=270]{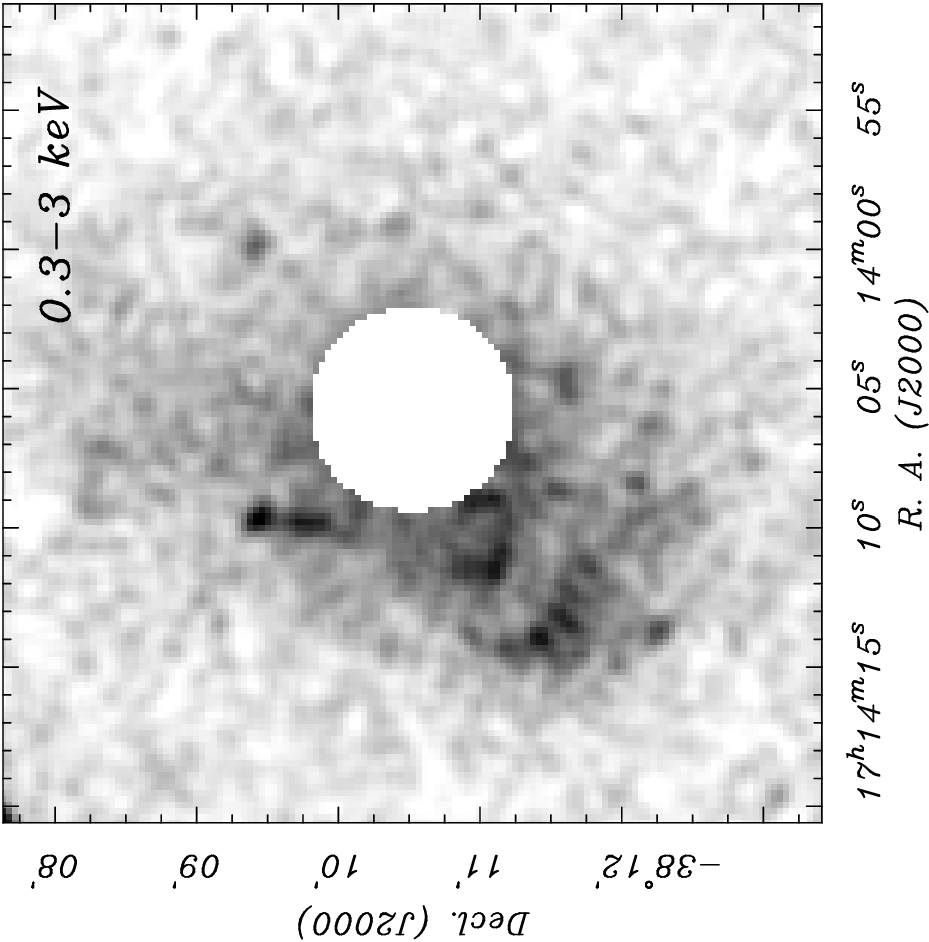}
\hfill
\includegraphics[height=0.48\linewidth,angle=270]{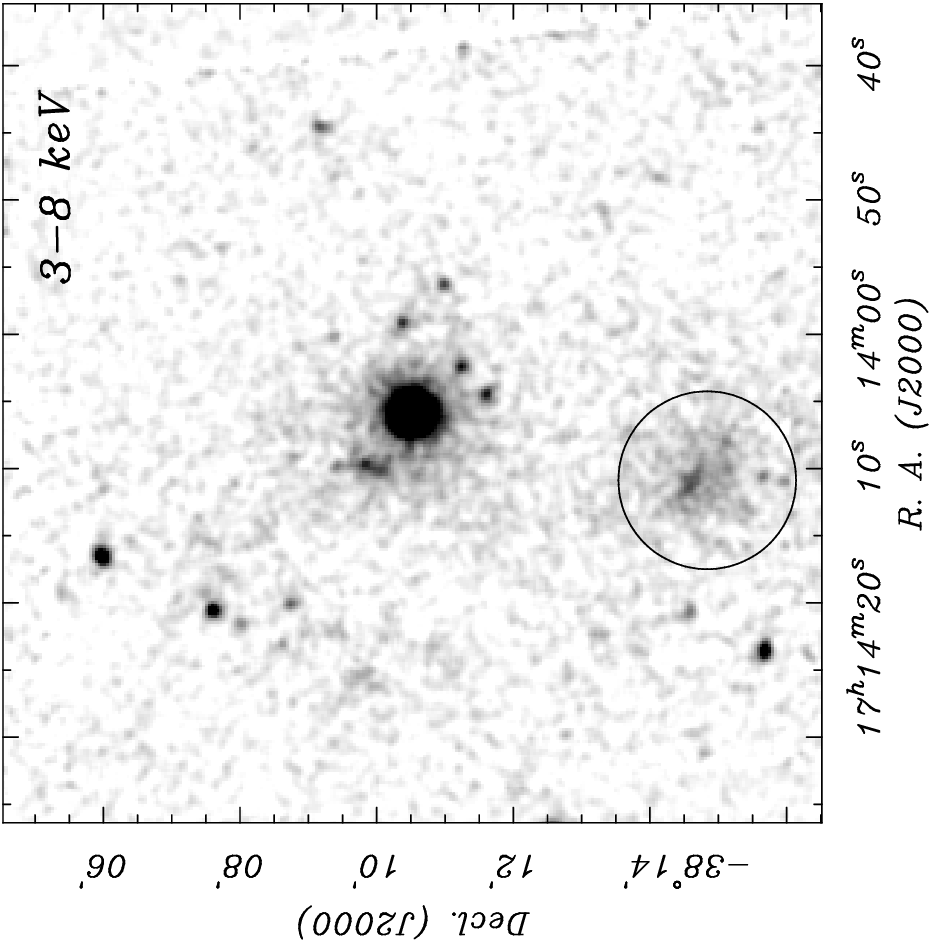}
}
\caption{ Exposure corrected, smoothed \xmm\ EPIC MOS
  images of \ccoshort\ in \ctb,
  in two energy bands: $0.3-3$~keV and $3-8$~keV.  These images are
  scaled logarithmically and stretched to highlight the diffuse emission.
  Diffuse thermal emission from the supernova remnant is
  clearly evident in the $0.3-3$~keV band, overlapping the radio
  remnant (contours).  The upper-left image shows the 20~cm contours
  from the Multi-Array Galactic Plane Imaging Survey \citep{whi05},
  while the upper-right image shows the 843~MHz contours from the
  Molonglo Observatory Synthesis Telescope \citep{gre99}.
  Details of X-ray structure of the remnant are shown with expanded
  scale in the lower-left image, with the magnetar region excluded.
  In the $3-8$~keV band (lower right), the circle to the south 
  south-east of the pulsar is the extraction aperture for the extended,
  hard source \blob\ analyzed in Section~\ref{sec:hard}.
  }
\label{fig:xmmimages}
\end{figure*}

\section {X-ray Observations}
\label{sec:obs}

\subsection {\nustar}

We observed \ccoshort\ twice with \nustar, on 2015 May~8 and 2016
September~22.  \nustar\ consists of two co-aligned X-ray telescopes, with
corresponding focal plane modules FPMA and FPMB that provide
$18^{\prime\prime}$ FWHM ($1^{\prime}$ HPD) imaging resolution over a
3$-$79~keV X-ray band, with a characteristic spectral resolution of
400~eV FWHM at 10~keV \citep{Harrison2013}.  The reconstructed
\nustar\ coordinates are accurate to $7\farcs5$ at the 90\% confidence
level.  The nominal timing accuracy of \nustar\ is $\sim$2~ms rms,
after correcting for drift of the on-board clock, with the absolute
timescale shown to be better than 3~ms \citep{Mori14, Madsen15}.

\nustar\ data were processed and analyzed using {\tt FTOOLS}
09May2016\_V6.19 ({\tt NUSTARDAS} 14Apr16\_V1.6.0) with \nustar\
Calibration Database (CALDB) files of 2016 July 6.  The resulting data
set provides a total of 80.9~ks and 78.6~ks of net good
exposure time for the 2015 and 2016 pointings, respectively. For all
subsequent analysis we merged data from both FPM detectors.

\subsection {\chandra}

Our previous \chandra\ observations of \ccoshort\ were reported in
Papers~I \& II (see Table~\ref{tab:logtable}). We acquired three
additional \chandra\ monitoring observations on 2012 July 16,
2015 May 4 (coordinated with our first \nustar\ observation),
and 2015 October 13.  The pulsar was located on the ACIS-S3 CCD and
recorded in continuous-clocking (CC) mode. This provided a time
resolution of 2.85~ms and no spectral pile-up.  This is the same
instrumental setup as used previously and fully described in the
earlier papers along with a description of their reduction and
analysis. The photon arrival times are adjusted in the standard
processing to account for the known position of the pulsar, spacecraft
dither, and any SIM offset. All observations were free of enhanced
background episodes and required no time filtering. Reduction and
analysis used the standard software packages CIAO (v4.8) and CALDB
(v4.1.1).

\subsection {\xmm}

\ccoshort\ was observed four times using \xmm, initially by Sato
\etal\ (2001) and later as part of our monitoring program (see
Table~\ref{tab:logtable}). In particular, our 2016 September 23 \xmm\
observation was obtained simultaneously with \nustar. In this work, we
concentrate on data from the three EPIC sensors on-board \xmm, the pn
detector \citep{Struder01} and MOS1 and MOS2 \citep{Turner01}. The pn
consist of 12 segments and the MOS comprises a mosaic of seven CCDs.
These detectors lie at the focal plane of coaligned replicated foil
mirrors with a maximum effective area of $A^{\rm eff}({\rm pn}) \approx
1400$~cm$^2$ and $A^{\rm eff}({\rm MOS1+MOS2}) \approx 1000$~cm$^2$ at
1.5~keV. The FWHM of the on-axis point spread function (PSF) at 1.5~keV is
$\approx12\farcs5$ and $\approx4\farcs3$, for the pn and MOS,
respectively. The EPIC detectors have a $\approx29^{\prime}$ diameter
field-of-view (FoV) and are sensitive to X-rays in the 0.15$-$12~keV
range with moderate energy resolution of $E/\Delta E({\rm pn}) \sim
$20$-$50.

The \xmm\ observations of \ccoshort\ were acquired with the EPIC~pn
operated in {\tt FullFrameMode} with a sampling time of 73.4~ms.  With
the exception of the first observation, data collected by the two EPIC
MOS detectors used the {\tt SmallWindowMode} for which only a small
portion ($1\farcm8 \times 1\farcm8$) of the central CCD is read out in
order to increase its time resolution to 0.3~s The first observation
(ObsID 0606020101) used {\tt FullFrameMode} mode that allowed MOS imaging
of the whole FoV at the nominal 2.6~s time resolution.  (The MOS
observations are not listed separately in Table~\ref{tab:logtable}.)

Data were reduced and analyzed using the Standard Analysis Software
(SAS) v.15 with the most up-to-date calibration files. After filtering
out background flares we obtained usable pn/MOS exposure time for the
observations, ordered by time, of 93/93~ks, 11.5/16.6~ks,
27.5/29.1~ks, and 20.0/21.6~ks. 

\section{Image Analysis}
\label{sec:image}

The \xmm\ data is most useful for mapping the supernova remnant flux
from \ctb. This soft thermal X-ray emissions falls outside the
\nustar\ energy band and the \chandra\ data contains only 1D spatial
information.  Figure~\ref{fig:xmmimages} displays EPIC MOS images in
two energy bands, above and below 3~keV.  These exposure-corrected,
smoothed images, combining data from all four \xmm\ data sets, are
scaled logarithmically and stretched to highlight the diffuse SNR
emission prominent in the lower energy band. The bulk of the thermal
emission lies in a region roughly $3^{\prime} \times 6^{\prime}$
in extent and orientated with a P.A. of $\sim$ $235^{\circ}$. To the east,
the emission is evidently delineated in part by the radio SNR shell
fragment. On a larger scale there is weak evidence for very low
surface-brightness, asymmetric X-ray emission,
possibly up to $\sim7^{\prime}$ from the magnetar, suggesting the true size
of the SNR.  Barely resolved, to the southeast, is a patch of radial
striations extending between $1\farcm7-2\farcm1$ that is not
associated with any known PSF pattern. The structure in this feature
is not resolved in the lower resolution EPIC~pn images (not shown).
Deep \chandra\ imaging-spectroscopy is required to further resolve and
identify the nature of these striations.

In the higher energy band, above 3~keV, the field is dominated by the
central magnetar with no clear evidence of SNR emission. Instead we
find a distinct diffuse hard X-ray feature $\approx2^{\prime}$ in diameter,
first reported by \cite{Nakamura09}. This source is located $4\farcm4$
south south-east of the magnetar with approximate coordinates (J2000.0)
R.A.$=17^{\rm h}14^{\rm m}10^{\rm s}\!.8$,
decl.$=-38^{\circ}14^{\prime}42^{\prime\prime}$
(herein \blob), and is not seen below $<3$~keV.
It is not clear whether this hard source is associated with \ctb. We
will explore its possible nature in Section~\ref{sec:hard}.

\begin{figure}[]
\includegraphics[width=0.97\linewidth,angle=0]{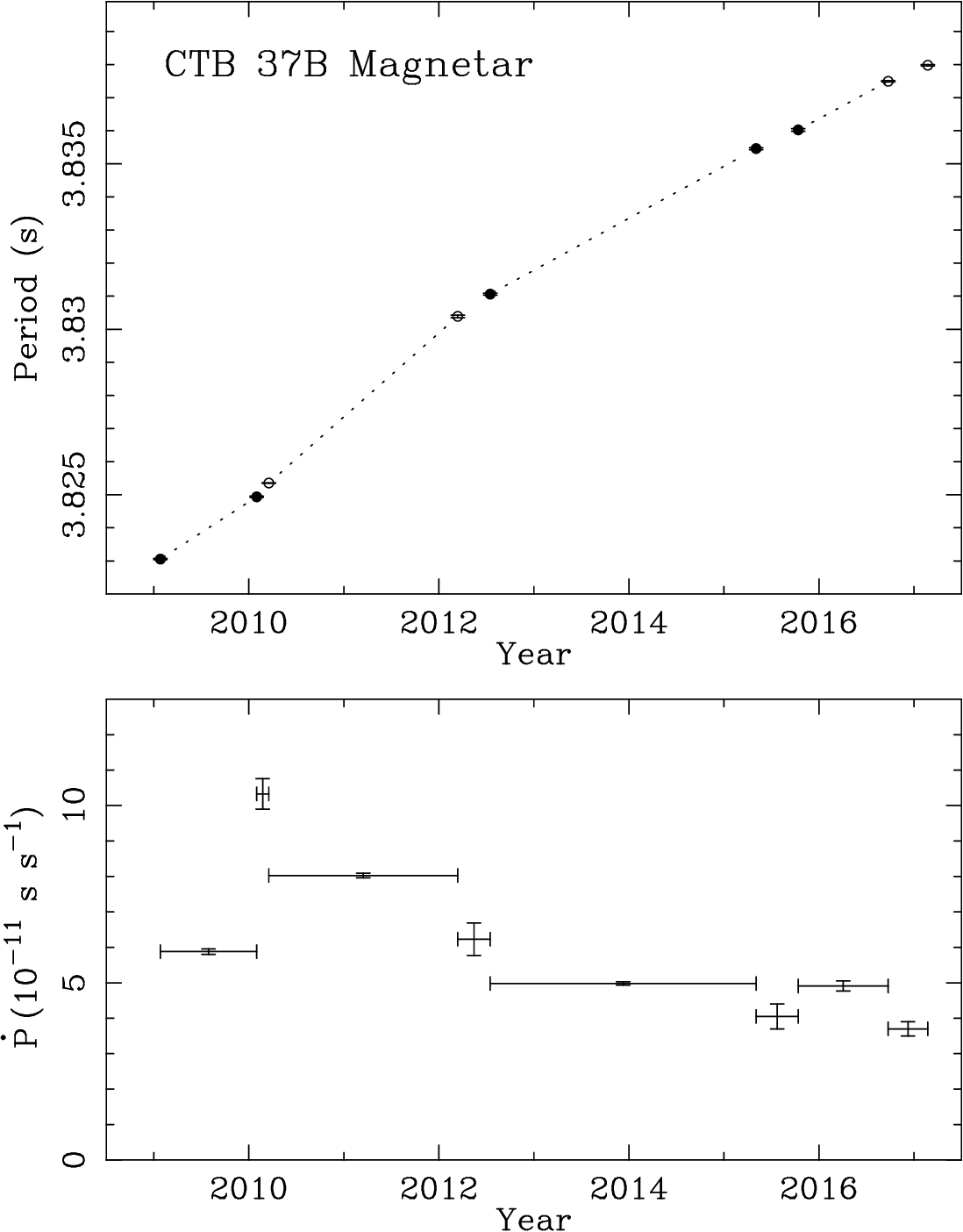}
\caption{
Top: Period measurements of \ccoshort\ from the five \chandra\ CC-mode observations
(filled circles) and the four \xmm\ observations.  Error bars are smaller than the size
of the symbols. Bottom: Period derivatives between consecutive measurements. The redundant 
\nustar\ points are not shown.
}
\label{fig:timing}
\end{figure}

\begin{figure}[t]
\includegraphics[height=0.97\linewidth,angle=270]{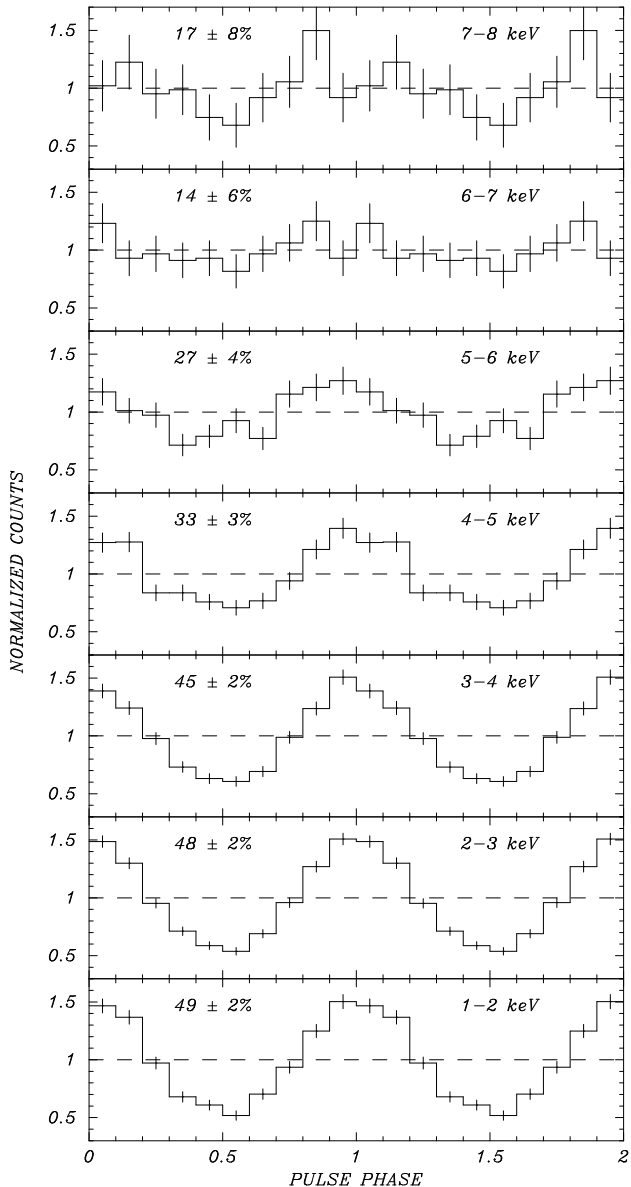}
\caption{ Pulse modulation of \ccoshort\ as a function of energy from
  the long, 2010 \xmm\ observation. These profiles are background
  subtracted and normalized so that the pulsed fraction can be read from
  the y-axis.  The pulsed signal decreases gradually with energy,
  approaching Gaussian fluctuations at 6$-$8 keV.  This is further
  illustrated in Figure~\ref{fig:modulation} for \nustar\ data.}
\label{fig:profiles}
\end{figure}

\begin{figure}[t] \hfill
\includegraphics[height=0.97\linewidth,angle=270]{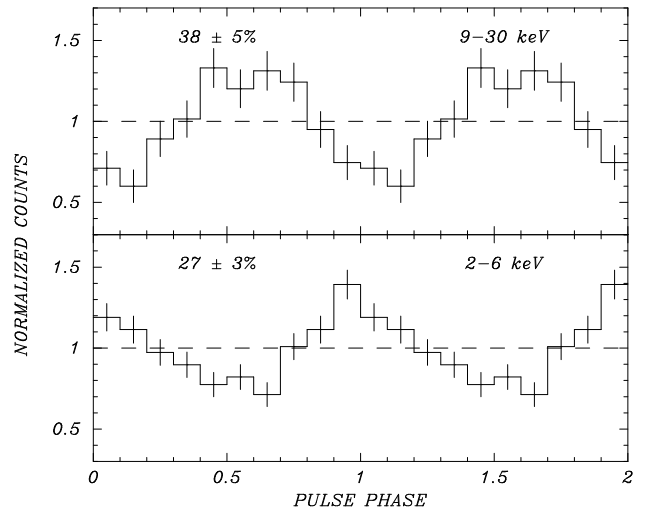}
\caption{\nustar\ pulse profiles of \ccoshort\ computed in two energy
  bands, 2$-$6~keV and 9$-$30~keV, from the 2016
  observation. These profiles are background subtracted and
  normalized so that the pulsed fraction is read from the y-axis. The
  $180^{\circ}$ shift in phase from the soft to the hard band is additional
  evidence for a hard spectral component that dominates above 9~keV
  (see discussion in Section~\ref{sec:model}). }
\label{fig:phaseshift}
\end{figure}

\section{Timing Analysis}
\label{sec:timing}

The erratic spin-down rates of magnetars is well documented, and
a similar behavior has already been established for \ccoshort\ (see
Paper~II).  Here we (re-)analyze all 8 years of accumulated timing
observations of this magnetar to characterize its long-term spin-down
evolution.  
For the following analysis, all photon arrival
times are converted to barycentric dynamical time (TDB) using the
DE405 solar system ephemeris and the \chandra\ position
(J2000.0) R.A.=$17^{\rm h}14^{\rm m}05^{\rm s}\!.74$,
decl.=$-38^{\circ}10^{\prime}30^{\prime\prime}\!.9$ (Paper~II).
The timing methods used herein are generally described in Paper~I.

Source lightcurves were generated for the \xmm\ data using a
$30^{\prime\prime}$ radius aperture in the energy range
1$-$5~keV, optimal for maximizing the pulse signal. As the \chandra\
data was obtained in CC-mode, we extracted counts from the central
four source columns ($2^{\prime\prime}$ diameter), with the $1-5$~keV energy
cut.  For \nustar, we extracted counts from a $51^{\prime\prime}$ radius
source aperture, including photons below the standard 3~keV spectral
analysis threshold.

To determine the spin period of \ccoshort\ at each epoch, we searched
around the nominal value using the $Z_1^2$ test \citep{buc83},
appropriate for the essentially sinusoidal pulse profile. The quoted
68.3\% ($1\sigma$) uncertainty in the period measurements are computed
from the $\Delta Z_1^2 (=\Delta \chi^2) = 1$ decrement resolved around
the signal peak in the periodogram.  The resulting period values are
reported in Table~\ref{tab:logtable} and are plotted in
Figure~\ref{fig:timing}, along with period derivatives, determined
by subtracting periods between each epoch.

In early 2010 the period derivative
quickly doubled, but it gradually returned to a more stable value of
$\dot P\approx5\times10^{-11}$.  Because of the sparse and irregular monitoring,
it is not possible to determine if additional, short episodes of enhanced
spin-down occurred.  There is one more pulsed detection,
from archival \asca\ data in 1996,
which gives $\dot P=7.0\times10^{-11}$
between 1996 and 2010 (Paper II), possibly representative
of a long-term, average value.
For clarity, we exclude the \asca\ point in Figure~\ref{fig:timing}.

{ The pulse profile appears to be generally stable in time and is
  well-modeled by a sinusoidal function $y(\phi) = A\sin\psi + B$ with
  pulsed fraction $f_p = A/B$.}  For the four \xmm\ epochs, the
average background-subtracted pulsed fraction is $\approx 44\pm4\%$ in
the 1$-$5~keV band. As shown in Figure~\ref{fig:profiles}, we also
examine the pulsed fraction as a function of energy using the deep
2010 \xmm\ observation. These data provide sufficient statistics to
resolve a strong energy dependence in the modulation, which decreases
with energy until reaching the $3\sigma$ limit of 18\% at $\approx
6$~keV.  This result is marginally reproduced in the other, much
shorter \xmm\ observations, but with less significance.

In the \nustar\ observations we at first found no evidence for a pulsed
signal in the full spectral band. Below 6~keV, however, we recover the
expected signal, consistent with the \xmm\ results. Examining the
modulation as a function of energy reveals a sinusoidal signal above
9~keV that is steadily increasing with energy,
but phase shifted by $\approx180\%$
relative to the softer X-ray pulse profile.  This is illustrated in
Figure~\ref{fig:phaseshift}, showing the pulse profiles in the two
energy bands of interest.  This strongly suggests the presence of
an independent hard spectral component dominating above 10~keV, 
consistent with that found for other magnetars, where the pulsed
fraction increases with energy.  However, the present \nustar\
data are insensitive to pulsations above 30~keV,
where the signal is most likely masked by a relatively large background.

\begin{deluxetable*}{lcccc}[!ht]
\tablewidth{0pt}
\tablecaption{\cco: Joint Fits \\ to Coincident 2016 \xmm /\nustar\ Spectra}
\tablehead{
\colhead{Parameter} & \colhead{CBB+PL} & \colhead{CBB+PL}  & \colhead{BB+PL}  & \colhead{PL+PL} \\
\colhead{}          & \colhead{}       & \colhead{Tuned}   & \colhead{}  & \colhead{}
} 
\startdata                                                  
$N_{\rm H}$~($10^{22}$ cm$^{-2}$)             & $4.0\pm0.5$         & $3.9\pm0.6$         & $3.6\pm0.5$           & $7.0\pm0.7$           \\
$kT_1$ (keV)                               & $0.54\pm0.05$       & $0.55\pm0.04$       & $0.62\pm0.04$         & \dots                 \\
$R_1$ (km)                                 & $1.3(1.1-1.6)$      & $1.3(1.1-1.4)$      & $1.1(0.95-1.3)$       & \dots                 \\
$L{_1}_{\rm bol}$\tablenotemark{a}           & \dots               & \dots               & $2.33 \times 10^{34}$  & \dots                 \\
$\Gamma_1$                                 & \dots               & \dots               & \dots                 & $3.94\pm0.4$          \\
$\Gamma_2$                                 & $0.59\pm0.22$       & $0.70\pm0.20$       & $0.92\pm0.26$         & $0.48\pm0.2$          \\
$\alpha$                                   & $1.6(1.2-2.6)$      & $2.1({\rm fixed}\tablenotemark{b})$         & \dots       & \dots   \\
$F_x(2-10\ {\rm keV})$\tablenotemark{c}    & $1.18\times10^{-12}$ & $1.18\times10^{-12}$ & $1.19\times10^{-12}$    & $1.15\times10^{-12}$   \\
$L_x(2-10\ {\rm keV})$\tablenotemark{a}    & $1.9\times10^{34}$   & $1.9\times10^{34}$   & $1.9\times10^{34}$     & $2.6\times10^{34}$     \\
$L_x(2-50\ {\rm keV})$\tablenotemark{a}    & $5.7\times10^{34}$   & $5.7\times10^{34}$   & $5.3\times10^{34}$     & $6.7\times10^{34}$     \\
$\chi^2_{\nu}(\nu)$                         & 0.869(69)           & 0.872(70)            & 1.02(70)              & 1.14(70)         
\enddata
\label{tab:coevalspec}                                                     
\tablecomments{\footnotesize Quoted uncertainties are at the 90\% confidence level for two interesting parameters.}
\tablenotetext{a}{\footnotesize Unabsorbed luminosity, in erg s$^{-1}$, for $d = 9.8$~kpc (Blumer \etal\ 2019).}
\tablenotetext{b}{\footnotesize Comptonized blackbody (CBB) parameter $\alpha \equiv -\ln(\tau_{es})/ \ln(A)$ tuned to match the modulation curve minimum. See Section~\ref{sec:model} for details.}
\tablenotetext{c}{\footnotesize Absorbed $2-10$~keV flux, in erg~cm$^{-2}$ s$^{-1}$, from the \xmm\ EPIC pn.}
\end{deluxetable*}

Given the strong energy dependence of the pulsed fraction, we postulate
a model for the modulation that incorporates the overlapping,
pulsed spectral components, including their phase difference.
The hard and soft components combine to cancel out the modulation in
their region of overlap.
We also model the modulation in detail in Section~\ref{sec:model}
to further discriminate among the acceptable spectral models.

\section{Spectral Analysis}
\label{sec:spectra}

Our previous analysis of \ccoshort\ showed that a variety of spectral
models can fit the 0.3$-$10~keV \chandra\ and \xmm\ data equally well,
including two blackbodies, a blackbody plus power-law, and a
Comptonized blackbody (Papers~I, II).  Now, the addition of the
\nustar\ data, covering 3$-$79~keV, adds a strong spectral lever arm
to help distinguish between these models. In the following sections,
we use the spectral and energy-dependent modulation of the pulsar to
show that its spectrum is best characterized by a Comptonized
blackbody whose temperature is constant in time,
plus an additional hard power-law.  We also analyze the spectrum of
the hard diffuse source \blob, located
$4\farcm4$ to the south of \ccoshort, to consider its nature.

For spectral fitting we use the {\tt XSPEC} (v12.8.2) package
\citep{Arnaud96} and characterize the column density with the built-in
{\tt TBabs} absorption model, selecting the {\tt wilm} Solar
abundances \citep{Wilms2000} and the {\tt vern} photoionization
cross-section \citep{Verner96}. The $\chi^2$ statistic is used to
evaluate the spectral fits throughout and the parameter uncertainties
are quoted at the 90\% confidence level for one or more
interesting parameters, as appropriate.  Response matrices and
ancillary response files were generated for each data set following
the standard procedures for their respective missions.

\subsection{\cco}
\label{sec:model}

In the current spectral study of \ccoshort\ we use all available
\chandra, \xmm, \nustar\ data sets\footnote{We excluded the 2007
  \chandra\ observation in TE mode as it provides a poor flux
  measurement.  The source was dithered on and off the edge of the CCD
  and suffers from time variable pile-up.}.  Source spectra
were extracted from circular apertures whose size were selected to
optimize the signal to noise ratio for each observation.  For the
\chandra\ CC-mode data we extracted source spectra from the sum of the
five central source columns, corresponding to a diameter of
$2.\!^{\prime\prime}5$ containing $\approx 95$\% of the point-source
enclosed energy. The background in this case was obtained from the
adjacent pixels on either side of the source region.  We estimate the
\nustar\ and \xmm\ backgrounds using an annular region, to allow for
underlying SNR emission in the source aperture. Because of the lack of
available background regions for the \xmm\ EPIC MOS data
acquired in {\tt SmallWindowMode}, only the EPIC pn data is used
for spectroscopy. Spectra from the two \nustar\ FPMs were co-added.

To characterize the broadband spectrum of \ccoshort\ we conducted a
joint fit of the contemporaneous 2016 \xmm\ and \nustar\ data.
Several trial spectral models were fitted in the 1$-$10~keV and
3$-$65~keV range, for the two missions, respectively. With the
addition of the harder \nustar\ spectra, we find that all single
component models are rejected, including those allowed by previous
fits to data below $10$~keV, as reported in Paper II.  Similarly, of
the plausible two-component models, the two-blackbody model is also
rejected due to a poor fit.

\begin{figure}[t]
\includegraphics[height=0.97\linewidth,angle=270]{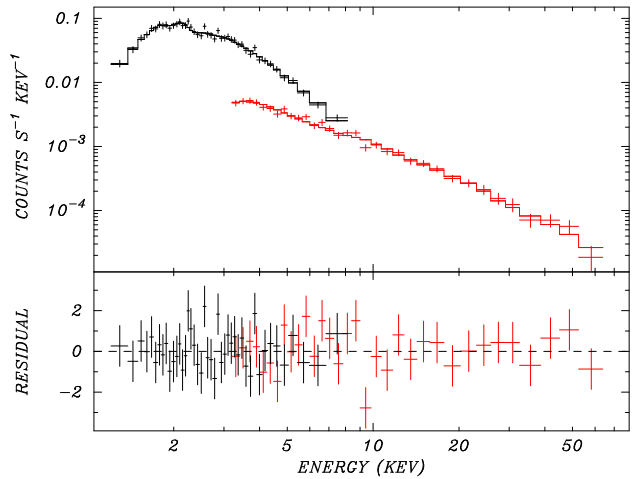}
\caption{ 
The broad-band X-ray spectrum of the magnetar \ccoshort\ in \ctb\
fitted to the absorbed Comptonized blackbody plus power-law model. Shown is the
joint fit to the coincident 2016 \xmm\ and \nustar\ data sets with model
normalizations allowed to be independent.  Upper panel: The data
points (crosses) are plotted along with the best fit model
(histogram) given in Table~\ref{tab:coevalspec}.  Lower panel: The
best fit residuals in units of sigma.
}
\label{fig:coevalspec}
\end{figure}

In contrast, we are able to obtain an excellent fit to the data
(Figure~\ref{fig:coevalspec}) using an absorbed blackbody plus hard
power-law model (BB+PL), with or without taking into account possible
Compton scattering of the thermal emission (CBB+PL). For the latter,
we use the model described in \cite{hal08}, where $\alpha \equiv
-\ln(\tau_{es})/\ln(A)$ is the log ratio of the scattering optical
depth $\tau_{es}$ over the mean amplification $A$ of photon energy per
scattering, valid for $\tau_{es}<<1$ \citep{ryb86}. The derived column
densities for these models are consistent with the value reported for
the SNR obtained using \suzaku\ data \citep{Nakamura09}.  While a fit
using the double power-law model is formally acceptable, in this case
the column density is far from that obtained for the SNR.  A summary
of spectral results for these models is presented in
Table~\ref{tab:coevalspec}.  Assuming a distance of 9.8~kpc
\citep{blu19}, the 2--50 keV X-ray luminosity of \ccoshort\ is
$\approx5.7\times10^{34}$ erg~s$^{-1}$.  This is comparable to its
spin-down power, $\dot E = 4\pi^2 I\dot P/P^3 \approx5\times10^{34}$
erg~s$^{-1}$ for $P=3.83$~s, $\dot P=7\times10^{-11}$, and moment of
inertia $I=10^{45}$ g~cm$^2$.

\begin{figure}[t]
\includegraphics[height=0.97\linewidth,angle=270]{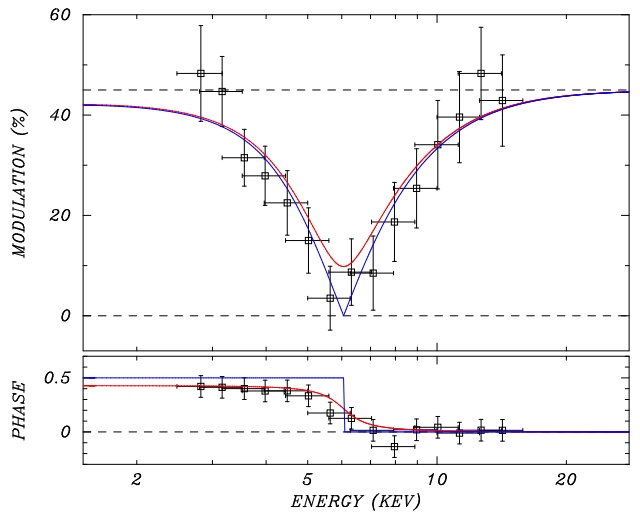}
\caption{
  Energy-dependent modulation $f_{p}(E)$ of \ccoshort\ computed using
  the 2016 \nustar\ observation. The modulation data points
  are derived from a sinusoidal fit to the background-subtracted
  pulse profiles, in overlapping logarithmically spaced energy bins.  The
  energy dependence of the modulation and phase (solid lines) are
  modeled well by the ratio of the spectral components of the
  Comptonized blackbody plus power-law model (see Section~\ref{sec:model}),
  with $f_1 = f_2 = 0.45$.
  The model is computed for the observed pulse-phase
  offset of $\Delta\phi = 0.43$ cycles between soft and hard X-rays (red line),
  and for a phase offset of $\Delta\phi = 0.5$ cycles (blue line).
  Other spectral models are rejected
  because the cross-over energies of their components do not
  match the energy (6.1~keV) where the observed pulsed fraction is a minimum. 
  }
\label{fig:modulation}
\end{figure}

A further constraint on possible spectral models is provided by the
energy-dependent modulation of the pulse profile as presented in
Section~\ref{sec:timing}.  For the sum of two sinusoidally varying
spectral components with the same period, but phase
difference $\Delta\phi$, the net pulsed fraction $f_{p}(E)$ as a
function of photon energy $E$ is predicted from their relative fluxes,
$F_1(E)$ and $F_2(E)$, as follows:

 $$ f_{p}(E) = {f_1 F_{1}(E)\sin[\psi(E)] + f_2 F_{2}(E)\sin[\psi(E) + \Delta\phi] \over
 F_{1}(E) + F_{2}(E)},  $$ 

\noindent where $f_1$ and $f_2$ are the (assumed energy-independent) pulsed
fractions of the two spectral components, and

$$\psi(E) = {\rm tan}^{-1}\left[{{{f_1 F_{1}(E) \over f_2 F_{2}(E)} + \cos(\Delta\phi)} \over \sin(\Delta\phi)} \right]$$

\noindent is the energy-dependent phase shift.  Of note, the
modulation tends towards a minimum at the spectral cross-over energy,
where the fluxes from the two spectral components are equal.

We apply this modulation model to the results of the joint fits to the
coincident 2016 \xmm\ and \nustar\ spectra for each spectral model of
Table~\ref{tab:coevalspec}. Only the Comptonized blackbody plus
power-law model is able to reproduce the observed modulation curve as
a function of energy, specifically the location of the dip at
$6.1$~keV (Figure~\ref{fig:modulation}). The implied
energy-independent modulation is 45\% for both components, and the
apparent phase offset is $\Delta\phi = 0.43$ cycles.  This offset
between the phase of the blackbody and the power-law spectral
components suggest that they arise on opposite sides of the neutron
star, or possibly that the thermal emission is not viewed directly,
but reflected by an opaque scattering screen (see
Section~\ref{sec:disc}).

The $\chi^2_{\nu}$ of the CBB+PL spectral fit is smaller than that of
the BB+PL model, justifying the extra parameter. But the modulation
data provides a further constraint on the spectrum that allows us to
distinguish between models, in this case implying deviations from pure
blackbody emission. Furthermore, simultaneously fitting the spectra
and modulation curve allows us to fine tune the spectral parameters
for the CBB+PL model to adjust the spectral cross-over (6.6~keV)
between components to match the dip in modulation at 6.1~keV.  The
tuned spectral parameters are well within the uncertainties of the
nominal fit parameters for this model, resulting in a negligible
change in the $\chi^2_{\nu}$, as presented in
Table~\ref{tab:coevalspec}.

\begin{figure}[]
\includegraphics[height=0.97\linewidth,angle=270]{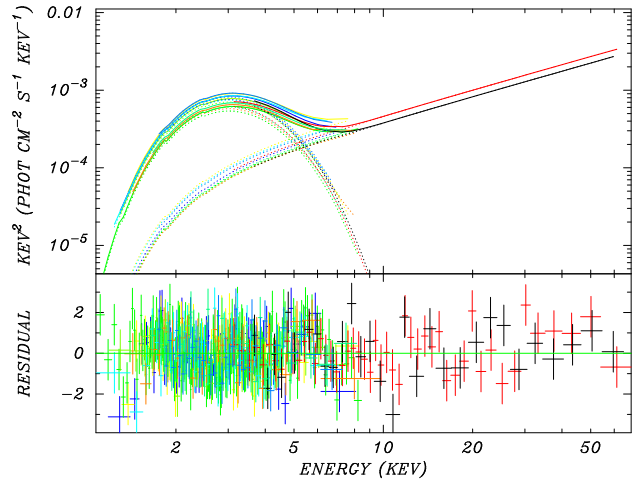}
\caption{{ Broad-band} X-ray spectra of the magnetar \ccoshort\ in \ctb\ at several
epochs from year 2009 onwards, as listed in Table~\ref{tab:logtable}.
Shown are \chandra\ ACIS, \xmm\ EPIC pn, and \nustar\ FPM spectra
fitted simultaneously to an absorbed blackbody plus power-law
model. The model normalizations are independent between spectra.
Upper panel: The data points (crosses) are plotted along with the best
fit model (histogram).  Lower panel: The best-fit residuals in units
of sigma.}
\label{fig:epochspectra}
\end{figure}

We now consider the full set of \chandra, \xmm, and \nustar\ spectra
acquired over a span of 8 years, { from 2009 January~25 to 2017
February~22 (11 observations, see Table~\ref{tab:logtable})}. Initial
fits at each epoch shows no evidence of significant change in the
spectral shape over time. We therefore fit all the spectra
simultaneously with their normalization left free, to allow for
calibration differences between telescopes and to search for flux
variability. For ease of comparison, we use the nominal blackbody plus
power-law model. The resulting combined fit is shown in
Figure~\ref{fig:epochspectra}.  The \nustar\ spectra, even if not
generally taken at the same epoch as the other data sets, strongly
constrain the temperature at the lower energies where the blackbody
component dominates the \chandra\ and \xmm\ spectra.  The combined
best fit model parameters are $N_{\rm H} = (3.88\pm0.16) \times
10^{22}$~cm$^{-2}$, blackbody temperature $kT =0.60\pm0.015$~keV, and
photon index $\Gamma_2 = 0.95\pm0.19$, with a $\chi^2_{\nu} = 1.02$
for 601 degrees of freedom. { This result is consistent with that
  presented in Table~\ref{tab:coevalspec} for the coincident \xmm\ and
  \nustar\ spectral fits.}

\begin{figure}[]
\includegraphics[height=0.97\linewidth,angle=270]{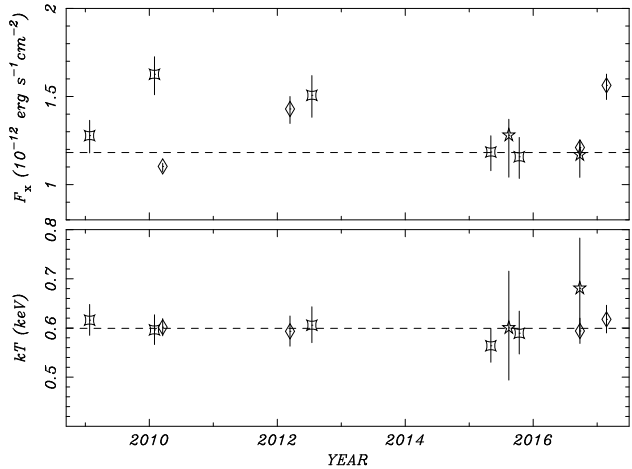}
\caption{ Time history of \ccoshort\ in \ctb. Plotted are the 2--10 keV
  fluxes (upper panel) and temperatures (lower panel) for
  \chandra\ (squares), \xmm\ (diamonds), and \nustar\ (stars)
  spectra presented in Figure~\ref{fig:epochspectra}. The plotted errors
  are for the 90\% confidence level.  }
\label{fig:spec_compare}
\end{figure}

To { study} the long-term flux and spectral variability we refitted
the individual spectra with column density and power-law index again
linked between epochs, but with temperatures and normalizations free.
The resulting temperature and flux for each observation are shown in
Figure~\ref{fig:spec_compare}. We find no significant change in the
time history of the temperature, but there is flux variability
evidently uncorrelated with temperature. We quantify this variability
by comparing a simultaneous fit across the 11 spectra with all
parameters linked to one for which the blackbody normalization (only)
is free to vary. The ratio of the resulting statistics, $\chi^2 =
2.30(621)$ and $\chi^2 = 1.04(610)$, strongly excludes a constant flux
model, with F-test probability $\wp < 1.5\times10^{-16}$.

We note that this result does not take into account systematic
differences in the inter-instrument flux calibrations or the
photometric reproducibility of the individual instruments.  However,
we find that observations which are contemporaneous or adjacent have
consistent fluxes, within their uncertainties.  Interestingly, a
possible exception is the highest flux point, in early 2010, which
coincides with the brief doubling of $\dot P$ in
Figure~\ref{fig:timing}. However, the data are otherwise too sparse to
test for a relation between luminosity and spin-down rate.


The \xmm\ and \chandra\ spectra used here were also analyzed by
\citet{wat19}, who found lower column densities, and 2--10~keV
fluxes higher by $\sim30\%$ than ours.  It is not clear what
is responsible for these differences.  Without the benefit
of the \nustar\ data, their spectral models do not account for
the hard power-law component.  They deemed the long-term flux
variability to be insignificant.

\subsection{\blob}
\label{sec:hard}

We extracted spectra from \blob, the hard source to the south of \ctb,
using data from the deep 2010 \xmm\ and the 2016 \nustar\
observations.  The source extraction aperture in the \xmm\ data
is shown in the lower right panel of Figure~\ref{fig:xmmimages}.
The 2015 \nustar\ observation is not used, as the source
region there is contaminated by stray light. For the \xmm\ data, we restrict
our analysis to the EPIC MOS data set due to the overwhelming
background in the EPIC pn. The MOS spectrum is fitted in the 2--7~keV band,
where the lower limit is chosen to avoid poorly subtracted, strong
instrumental features. The \nustar\ spectrum is fitted
in the 3--20~keV range, for lack of photons at higher energies.  A
total of 2310 MOS counts and 1010 FPM counts are extracted from the
$2^{\prime}$ diameter aperture, of which 45\% and 65\% are background
counts, respectively. The background regions for both spectra are
chosen from a $4^{\prime}$ diameter aperture adjacent to the source region.
With the lack of evidence for spectral features, we fit a simple
absorbed power-law model. The best-fit parameters are $N_{\rm H} =(11\pm4)
\times 10^{22}$~cm$^{-2}$, and photon index $\Gamma = 2.2^{+0.6}_{-0.5}$,
with a $\chi^2_{\nu} = 0.62$ for 49 degrees of freedom. These results
are somewhat different from the findings of \cite{Nakamura09} from
\suzaku, particularly our 2--10~keV unabsorbed flux of
$2.5 \times 10^{-13}$ erg\,cm$^{-2}$\,s$^{-1}$, which is half that reported by those authors.
We consider the present results to be more reliable because of the
smaller PSF of \nustar\ and \xmm\ in comparison to the \suzaku\ data,
which suffered contamination from the nearby SNR and magnetar.
The fitted column density to \blob\ is significantly higher than
in the favored spectral model for \ccoshort, which suggests that it might be
unrelated to the magnetar or the SNR.

\begin{figure}[]
\includegraphics[height=0.97\linewidth,angle=270]{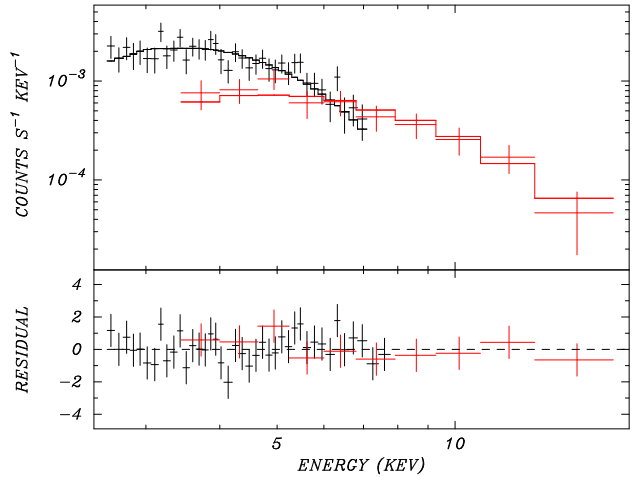}
\caption{The broad-band X-ray spectrum of \cco, the hard diffuse
   source south of \ctb, plotted along with the best fit absorbed
   power-law model (histogram) given in the text. The lower panel
   displays the residuals from the best fit model in units of sigma.  }
\label{fig:blobspec}
\end{figure}

\section{Discussion and Conclusions}
\label{sec:disc}

{ The spectra of most magnetars are well-fitted below 10~keV with
  either a two-blackbody model or a blackbody plus a non-thermal
  power-law model with a steep $\Gamma\approx4$ slope.  The latter
  model is similar to a Comptonized blackbody of appropriate
  scattering parameters. Evident non-thermal emission above 10~keV is
  characterized by a much flatter power-law with $\Gamma \simlt 1.5$.
  The spectrum of \ccoshort\ is somewhat unusual in that the flux
  below $10$~keV falls off much slower with energy then expected.  In
  addition to a thermal component of $kT \approx 0.6$, notably, a hot
  thermal component ($kT \approx 2.8$~keV) or a flat ($\Gamma \approx
  0.8$) power-law component is required to fit the spectrum in this
  band.  Because its power-law index is compatible with those from
  fits obtained above 10~keV, only two components are in fact required
  to fit the overall broad band spectrum of \ccoshort.}

For the magnetar spectral survey of \citet{eno17}, only four out of 15
objects that have a hard component above 10~keV can be satisfactorily
fit with a single component in addition to their thermal spectra:
SGR~1806$-$20, SGR~1900+14, 1E~1547.0$-$5408, and SGR~1833$-$0832.
Furthermore, two more objects, 1E~1048.1$-$5937, and 1E~2259+586, show
the need for either one or two components at different times,
suggesting a correlation with their emission state.  { Like for
  SGR~1806$-$20 and SGR~1900+14, the hard power-law emission from
  \ccoshort\ might dominate and obscure a fainter thermal/power-law
  component.}  The spectral properties of \ccoshort, shared by several
of the SGRs, may therefore be attributes of its comparative youth and
greater magnetic field strength relative to other AXPs, as shown in
Figure~\ref{fig:comp}.
{ In fact, the spin-down rate, characteristic age, and surface
  magnetic field of \ccoshort\ are most similar to those of
  SGR~1900$+$14.  And its hardness ratio, comparing the flux in the
  hard to soft bands, of $F_{15-60~{\rm kev}}/F_{1-10~{\rm keV}} =
  2.51$ agrees well with the correlations for those parameters reported
  by \cite{eno17}.
}


The energy-dependent modulation of \ccoshort\ is also unusual for a
magnetar.  In contrast with most magnetars, its pulsed fraction
decreases with energy up to $\approx6$~keV, but this can be explained
by the rising contribution of the hard spectral component, which
happens to be phase-shifted by $\approx 0.43$ cycles from the soft
X-rays, thus reducing the net modulation.  A similar effect is seen in
1RXS J170849.0$-$400910 \citep{den08}, where a hard pulse component
that dominates above 8~keV is shifted by $\approx 0.4$ cycles from a
soft pulse component below that energy.

The simplest interpretation of the phase shift would have the two
spectral components arise from nearly opposite sides of the NS.
However, in the \citet{bel13a,bel13b} model of scattering from
e$^{\pm}$ pairs on twisted magnetic field-line bundles, the particles
can form an opaque (to soft X-rays) resonance-scattering layer at the
top of the magnetic loops, which obscures the surface hot spots from
certain viewing angles.  In this case, the thermal emission might not
be seen directly, but could peak in reflection at the opposite phase.

For most other magnetars with hard pulsed components, the
modulation increases with energy above 10~keV. Unfortunately
the statistics of the \nustar\ data on \ccoshort\ are not sufficient
above 30~keV to determine if its pulsed modulation continues
to increase with photon energy above 10~keV, where the pulsed
fraction reaches $\approx0.45$.  In some objects the pulsed
fraction increases to as high as 0.96 at 20~keV
(1E 2259+586:\citealt{vog14}), while in others, the
pulsed fraction levels off 
(4U 0142+61: \citealt{ten15}; 1E1841$-$045: \citealt{an13, an15})
or decreases
(1E1048.1-5937: \citealt{yan16}; XTE J1810$-$197: \citealt{got19})
at high energy.

Our original motivation for monitoring \ccoshort\
was to detect any timing and spectral changes 
that might {\it precede} an SGR outburst, which could
reveal the proximate cause of SGR flares. 
Based on the similarity of the timing properties of
\ccoshort\ to the most energetic SGRs, an
outburst can be expected, although we are still
waiting for it.  The factor of 2 change in $\dot P$
that we observed is evidently not sufficient to trigger
an outburst, even though it may be correlated with
changes in X-ray luminosity.   { In other magnetars, an
increase in torque has been seen {\it following} an
outburst, by $\sim 100$ days in 1E~1048.1$-$5937 \citep{arc15},
and by $\sim1$~year in SGR~1820$-$20 \citep{woo07,you15},
so it cannot be excluded that \ccoshort\ had an
undetected bursting episode just prior to the beginning
of our timing program.}

The relation, if any, between the magnetar and \tev\ remains
uncertain.  There is no direct evidence from X-rays
for a PWN that could contribute TeV emission.
The existing \xmm\ images are
limited in their sensitivity to a PWN
by the thermal X-ray emission from the SNR
and the magnetar itself, while the deep imaging with \chandra\ 
that would be necessary to reveal a
compact, nonthermal structure, has not yet been done.
As discussed in Paper II, even though the {\it present\/}
spin-down power of \ccoshort\ is only about equal
to the TeV luminosity of \tev, it is plausible that
relic electrons, from a recent time when the $\dot E$ was
much larger than it is now, are powering a TeV nebula
via inverse Compton scattering.
GeV emission from {\it Fermi} has also been
detected in the direction of \snr\ \citep{xin16},
and its spectrum connects smoothly with that of \tev,
suggesting that there is only one $\gamma$-ray source
at this location.

{ It is interesting that the only magnetar that has good evidence
of possessing an X-ray PWN, Swift J1834.9$-$0846 \citep{you16},
is also coincident with a SNR (W41), a TeV source (HESS J1834$-$097),
and a GeV source (2FGL J1834.3$-$0848). \citet{abr15} discuss
these associations in the context of the PWN possibly
belonging to the nearby pulsar candidate XMMU J183435.3$-$084443
\citep{muk09}, but it more likely belongs to Swift J1834.9$-$0846
(see also \citealt{mis11}).
\ccoshort\ thus bears some similarity to Swift J1834.9$-$0846
in its associations.}

The nearby source \blob\ is also a potential $\gamma$-ray
emitter, although it is not clear whether it is part of
the SNR shell of \snr\ or an unrelated object.  Its
X-ray $N_{\rm H}$, larger than that of the magnetar or the
SNR, is tentative evidence that it may be a background source,
e.g., a PWN. The GeV and TeV centroids are both closer
to the magnetar and the SNR than to \blob,
although there is room for overlap.  This leaves the
nature of \blob\ unknown.  It could also be investigated
with a deeper \chandra\ observation, e.g., to search for
an embedded pulsar point source.
 
\acknowledgements
We thank the referee for making several informative suggestions.
This investigation  is based on new data obtained with the \nustar\
and \xmm\ Observatories, and on archival data obtained with the \chandra\
Observatory.  Support for this work was provided by NASA through
\nustar\ Cycle~2 Guest Observer Program grant NNX17AC22G and through
\xmm\ Cycle~15 Guest Observer Program grant NNX17AC14G.  The \nustar\
mission is a project led by the California Institute of Technology,
managed by the Jet Propulsion Laboratory, and funded by the National
Aeronautics and Space Administration.  This research made use of the
\nustar\ Data Analysis Software (NuSTARDAS) jointly developed by the
ASI Science Data Center (ASDC, Italy) and the California Institute of
Technology (USA). \xmm\ is an ESA science mission with instruments and
contributions directly funded by ESA member states and NASA.  This
research has made use of data and software provided by the \chandra\
X-ray Center (CXC) that is operated for NASA by the Smithsonian
Astrophysical Observatory. 
Data and software were also provided by the
High Energy Astrophysics Science Archive Research Center (HEASARC),
which is a service of the Astrophysics Science Division at NASA/GSFC
and the High Energy Astrophysics Division of the Smithsonian
Astrophysical Observatory.  We also acknowledge extensive use of the
arXiv and the NASA Astrophysics Data Service (ADS). E.V.G. thanks
Josep Maria Paredes for hosting his sabbatical at the University of
Barcelona Institut de Ci\`encies del Cosmos (ICCUB) and acknowledges
support through the ``Programa Estatal de Foment de la Investigaci\`o
Cient\'{\i}fica i T\`ecnica d'Excell\`encia, Convocat\`oria 2014,
Unitats d'Excell\`encia {\it Maria de Maeztu}.''

\smallskip
Facilities: \textit{NuSTAR, CXO, XMM}


\newpage

\begin{thebibliography}{}

\bibitem[Abramowski \etal(2015)]{abr15}
{ Abramowski, A., Aharonian, F., Ait Benkhali, F., \etal\ 2015, \aap, 574, A27}

\bibitem[Aharonian \etal(2006)]{aha06}
Aharonian, F., Akhperjanian, A. G., Bazer-Bachi, A. R.,
\etal\ 2006, \apj, 636, 777

\bibitem[Aharonian \etal(2008)]{aha08a}
Aharonian, F., Akhperjanian, A. G., Barres de Almeida, U.,
\etal\ 2008, \aap, 486, 829

\bibitem[An \etal(2013)]{an13}
An, H., Hasco\"et, R., Kaspi, V. M., et al. 2013, \apj, 779, 163

\bibitem[An \etal(2014)]{an14}
An, H., Kaspi, V. M., Beloborodov, A.~M., et al. 2014, \apj, 790, 60

\bibitem[An \etal(2015)]{an15}
An, H., Archibald, R. F., Hasco\"et, R., Kaspi, V. M., et al. 2015,
\apj, 807, 93

\bibitem[Archibald \etal(2015)]{arc15}
{ Archibald, R. F., Kaspi, V. M., Ng, C.-Y., \etal\ 2015, 800, 33}

\bibitem[Arnaud(1996)]{Arnaud96} Arnaud, K. A. 1996, in ASP Conf. Ser. 101,
Astronomical Data Analysis Software and Systems V, ed. G. H. Jacoby
\& J. Barnes (San Francisco, CA: ASP), 17

\bibitem[Beloborodov(2013a)]{bel13a}
Beloborodov, A. M. 2013a, \apj, 762, 13

\bibitem[Beloborodov(2013b)]{bel13b}
Beloborodov, A. M. 2013b, \apj, 777, 114

\bibitem[Blumer \etal(2019)]{blu19}
Blumer, H., Safi-Harb, S. \& Kothes, R. 2019, \mnras, submitted
(arXiv:1906.07249)

\bibitem[Buccheri \etal(1983)]{buc83}
Buccheri, R., Bennet, K., Bignami, G. F., \etal\ 1983, \aap, 128, 245

\bibitem[Camilo \etal(2018)]{cam18}
Camilo, F., Scholz, P., Serylak, M., \etal\ 2018, \apj, 856, 108

\bibitem[den Hartog \etal(2004)]{den04}
den Hartog, P. R., Kuiper, L., Hermsen, W., \& Vink, J. 2004, ATel, 293, 1

\bibitem[den Hartog \etal(2004)]{den08}
den Hartog, P. R., Kuiper, L., \& Hermsen, W. 2008, \aap, 489, 263

\bibitem[Enoto \etal(2017)]{eno17}
Enoto, T., Shibata, S., Kitaguchi, T., \etal\ 2017, ApJS, 231, 847, L25

\bibitem[Gotthelf \etal(2019)]{got19}
Gotthelf, E. V., Halpern, J. P., Alford, J. A. J., \etal\ 2019, \apj, 974, L25

\bibitem[Green \etal(1999)]{gre99}
Green, A. J., Cram, L. E., Large, M. I., \& Ye, T. 1999, \apjs, 122, 207

\bibitem[Halpern \& Gotthelf(2010a)]{hal10a}
Halpern, J. P., \& Gotthelf, E. V. 2010a, \apj, 710, 941 (Paper~I)

\bibitem[Halpern \& Gotthelf(2010b)]{hal10b}
Halpern, J. P., \& Gotthelf, E. V. 2010b, \apj, 725, 1384  (Paper~II)

\bibitem[Halpern et al.(2008)]{hal08}
Halpern, J. P., Gotthelf, E. V., Reynolds, J., Ransom, S. M. \& Camilo, F.
2008, \apj, 676, 1178

\bibitem[Harrison et al.(2013)]{Harrison2013} 
Harrison, F.~A., Craig, W.~W., Christensen, F.~E., et al. 2013,
\apj, 770, 103   

\bibitem[Hasco\"et et al.(2014)]{has14}
Hasco\"et, R., Beloborodov, A. M., \& den Hartog, P. R. 2915, \apj, 786, L1

\bibitem[Kaspi \etal(2014)]{kas14}
Kaspi, V. M., Archibald, R. F., Bhalerao, V., \etal\ 2014, \apj, 786, 84

\bibitem[Madsen et al.(2015)]{Madsen15} 
Madsen, K.~K., Harrison, F.~A., Markwardt, C.~B., et al. 2015, ApJS, 220, 8

\bibitem[Mereghetti \& Stella(1995)]{mer95}
Mereghetti, S., \& Stella, L. 1995, \apj, 442, L17

\bibitem[Misanovic \etal(2011)]{mis11}
{ Misanovic, Z., Kargaltsev, O., \& Pavlov, G. G. 2011, \apj, 735, 33}

\bibitem[Molkov \etal(2004)]{mol04}
Molkov, S. V., Cherepashchuk, A. M., Lutovinov, A. A.,
\etal\ 2004, AstL, 30, 534

\bibitem[Mori et al.(2013)]{mor13} 
Mori, K., Gotthelf, E.~V., Zhang, S., et al. 2013, \apj, 770, L23

\bibitem[Mori et al.(2014)]{Mori14} 
Mori, K., Gotthelf, E.~V., Dufour, F., et al. 2014, \apj, 793, 88

\bibitem[Mukherjee \etal(2009)]{muk09}
{ Mukherjee, R., Gotthelf, E. V., Halpern, J. P., \etal\ 2009, \apj, 691, 1707}

\bibitem[Nakamura \etal(2009)]{Nakamura09}
Nakamura, R., Bamba, A., Ishida, et al. 2009, \pasj, 61, S197

\bibitem[Revnivtsev \etal(2004)]{rev04}
Revnivtsev, M. G., Sunyaev, R. A., Varshalovich, D. A.,
\etal\ 2004, Astl, 30, 382

\bibitem[Rybicki \& Lightman(1986)]{ryb86} Rybicki, G. B., \& Lightman, A. P.
1986, Radiative Processes in Astrophysics, (New York: Wiley-VCH)

\bibitem[Sato \etal(2010)]{sat10}
Sato, T., Bamba, A., Nakamura, R., \& Ishida, M. 2010,
\pasj, 62, L33

\bibitem[Str\"uder \etal(2001)]{Struder01} 
Str\"uder, L., Briel, U., Dennerl, K., \etal\ 2001, A\&A, 365, L18

\bibitem[Tamba \etal(2019)]{tam19}
Tamba, T., Bamba, A., Odaka, H., \& Enoto, T. 2019, \pasj, in press
(arXiv:1906.04406)
 
\bibitem[Tendulkar \etal(2015)]{ten15}
Tendulkar, S. P., Hasco\"et, R., Yang, C., \etal\ 2015, \apj, 808, 32

\bibitem[Turner \etal(2001)]{Turner01}
Turner, M. J. L., Abbey, A., Arnaud, M., et al. 2001, \aap, 365, L27

\bibitem[van Paradijs \etal(1995)]{van95}
van Paradijs, J., Taam, R. E., \& van den Heuvel, E. P. J. 1995, \aap, 299, L41

\bibitem[Verner et al.(1996)]{Verner96} 
Verner, D.~A., Ferland, G.~J., Korista, K.~T., \& 
Yakovlev, D.~G. 1996, \apj, 465, 487

\bibitem[Vogel \etal(2014)]{vog14}
Vogel, J. K., Hasco\"et, R., Kaspi, V. M., \etal\ 2014, \apj, 789, 75

\bibitem[Watanabe \etal(2019)]{wat19}
Watanabe, H., Bamba, A., Shibata, S., \& Watanabe, E. 2019, \pasj, in press
(arXiv:1905.11561)

\bibitem[White \etal(2005)]{whi05}
White, R. L., Becker, R. H., \& Helfand, D. J. 2005, \aj, 130, 586

\bibitem[Wilms et al.(2000)]{Wilms2000} 
Wilms, J., Allen, A., \& McCray, R. 2000, \apj, 542, 914

\bibitem[Woods et al.(2007)]{woo07}
{ Woods, P. M., Kouveliotou, C., Finger, M. H. \etal\ 2007, \apj, 654, 470}

\bibitem[Xin \etal(2016)]{xin16}
Xin, Y.-L., Liang, Y.-F., Li, X., \etal\ 2016, \apj, 817, 64

\bibitem[Yang \etal(2016)]{yan16}
Yang, C., Archibald, R. F., Vogel, J. K., \etal\ 2016, \apj, 831, 80

\bibitem[Younes \etal(2015)]{you15} 
{ Younes, G., Kouveliotou, C., \& Kaspi, V. M. 2015 \apj, 809, 165}

\bibitem[Younes \etal(2016)]{you16} 
{ Younes, G., Kouveliotou, C., Kargaltsev, O., \etal\ 2016 \apj, 824, 138}

\bibitem[Younes \etal(2017a)]{you17a} 
{ Younes, G., Baring, M. G., Kouveliotou, C., \etal\ 2017a \apj, 851, 17 }

\bibitem[Younes \etal(2017b)]{you17b} 
{ Younes, G,, Kouveliotou, C., Jaodand, A., \etal\ 2017b \apj, 847, 85}

\end{thebibliography}
\end{document}